\documentclass[twocolumn]{emulateapj}
\usepackage{graphicx}
\usepackage{natbib}

\begin{document}

\title{Why are dense planetary  rings only found between 8 and 20 AU?}
\author{M. M. Hedman}
\affil{Physics Department, University of Idaho,  875 Perimeter Drive, MS. 0903, Moscow ID 83844-0903 \\ {\tt mhedman@uidaho.edu}}

\begin{abstract}
The recent discovery of dense rings around the Centaur Chariklo (and possibly Chiron) reveals that complete dense planetary rings are not only found around Saturn and Uranus, but also around small bodies orbiting in the vicinity of those giant planets. This report examines whether there could be a physical process that would make rings more likely to form or persist in this particular part of the outer Solar System. Specifically, the ring material orbiting Saturn and Uranus appears to be much weaker than the material forming the innermost moons of Jupiter and Neptune. Also, the mean surface temperatures of Saturn's, Uranus' and Chariklo's rings are all close to 70 K. Thus the restricted distribution of dense rings in our Solar System may arise because icy materials are particularly weak around that temperature. 
\end{abstract}

\maketitle

\section{Introduction}

The recent discovery of rings around  Chariklo \citep{BR14} not only demonstrates that small bodies can have substantial rings, but also hints that ringed worlds might not be randomly distributed within our Solar System. Chariklo's orbit keeps it between 13 AU and 19 AU from the Sun, so Chariklo is always located between the orbits of Uranus and Saturn, the two giant planets with the densest and most extensive ring systems. Furthermore, a recent occultation of Chiron, an object orbiting through roughly the same region of space (between 8 and 19 AU),  revealed narrow dips that might represent rings similar to Chariklo's \citep{Ruprecht15, Ortiz15}. Thus there may be as many as four worlds with high-opacity rings between 8 and 20 AU from the Sun. By contrast, there is no evidence for similarly complete rings with comparably high optical depths outside this region. It has long been known that Jupiter has the least substantial ring system of any giant planet \citep{Burns04} and that the only features in Neptune's rings with substantial opacity are the arcs in the Adams ring \citep{Porco95} whose distribution and intensity has varied dramatically over the years \citep{dePater05}. With the discovery of rings around Chariklo and possibly Chiron, it is also remarkable that no one has reported evidence for dense rings around any Near Earth Objects, Main Belt Asteroids or Kuiper Belt Objects, despite numerous radar observations,  spacecraft encounters and high-quality stellar occultations that could have revealed such features. 

Of course, the lack of dense, complete rings around Jupiter and Neptune could simply reflect the diverse histories of giant planets, and future surveys may eventually find rings around some asteroids or Kuiper Belt Objects. However, if the observed distribution of dense rings is not just an artifact of small number statistics and incomplete data, then this might have implications for both the architectures of the giant planet systems and the  distribution of rings orbiting extrasolar planets \citep{BF04, SC11, Mamajek12}. Hence it is worth considering the possibility that there  is some physical process that favors the formation or maintenance of dense rings 8-20 AU from the Sun. 

This report briefly explores a few basic aspects of the known dense planetary rings that could potentially help explain their distribution. First, the overall architectures of the ring-moon systems surrounding the giant planets indicate that the ring material around Saturn and Uranus is much weaker than the material in the moons  orbiting  Jupiter and Neptune.  At the same time, the trends in the albedos of the known dense rings suggest that the rings around Saturn, Uranus and Chariklo have similar temperatures ($\sim$70 K). Hence the restricted distribution of dense rings in our Solar System might be explained if ice-rich material becomes particularly weak at temperatures close to 70 K, and if such temperatures can only be achieved between 8 AU and 20 AU from the Sun.

\section{The material in dense rings is very weak}

While there are no direct measurements of the strength of any objects in the outer Solar System,  an examination of the basic architectures of the ring-moon systems surrounding each giant planet suggests that the materials in Saturn's and Uranus' rings are much weaker than the material in orbit around Jupiter and Neptune. Such considerations also indicate that the solid material in orbit around either a giant planet or a smaller body probably needs to be very weak to form dense rings. 

The Saturn and Uranus systems have similar basic structures, with dense rings close to the planet, moons further out,  and a relatively narrow zone in between where moons and rings coexist. These basic aspects of the Saturn and Uranus systems can be most easily explained by assuming the relevant solid matter has negligible internal strength and identifying the transition zone between rings and moons with the Roche Limit:
\begin{equation}
a_{R}=\left(\frac{3M_p}{\gamma \rho}\right)^{1/3},
\end{equation}
where $M_p$ is the mass of the central planet, $\rho$ is the orbiting matter's mass density, and  $\gamma$ is a numerical coefficient that depends on the assumed shape and spin state of the solid objects \citep{Tiscareno13}. Exterior to $a_{R}$, material can aggregate under its own gravity to form moons, but inside $a_{R}$ the material cannot aggregate and so forms rings. This model not only provides a useful qualitative picture of these systems, it also yields a reasonable estimate of the typical mass density for the solid material orbiting around Saturn and Uranus. Placing $a_{R}$ at the outer edge of Saturn's main rings requires that $\rho = \gamma^{-1}0.7$ g/cm$^3$. Assuming $\gamma \simeq 1.6$ (which is reasonable for objects that nearly fill their Roche Lobes, see Tiscareno et al. 2013), this implies that $\rho \simeq 0.4$ g/cm$^3$.  Such a mass density is comparable to those of Saturn's smaller moons \citep{Thomas13}, and implies that the material in Saturn's rings and inner small moons has a substantial porosity. Similarly, if $a_{R}$ is close to Uranus' outermost dense ring (the epsilon ring), then $\rho = \gamma^{-1}1.9$ g/cm$^3$ and $\rho \simeq 1.2$ g/cm$^3$, which is not an unreasonable mass density for (possibly porous) carbon-rich materials in the Uranus system \citep{Tiscareno13}. 

This simple model of essentially strengthless material does not work for the ring-moon systems of Jupiter and Neptune.  For example, Neptune's  most opaque ring (the Adams ring) lies substantially outside the orbits of three moons, Naiad, Thalassa and Despina, so there is no clear transition zone between rings and moons that can be interpreted as a particular Roche limit. Furthermore, the locations of both Jupiter's and Neptune's innermost moons suggest that these objects have finite strength.

\begin{figure}
\resizebox{3.2in}{!}{\includegraphics{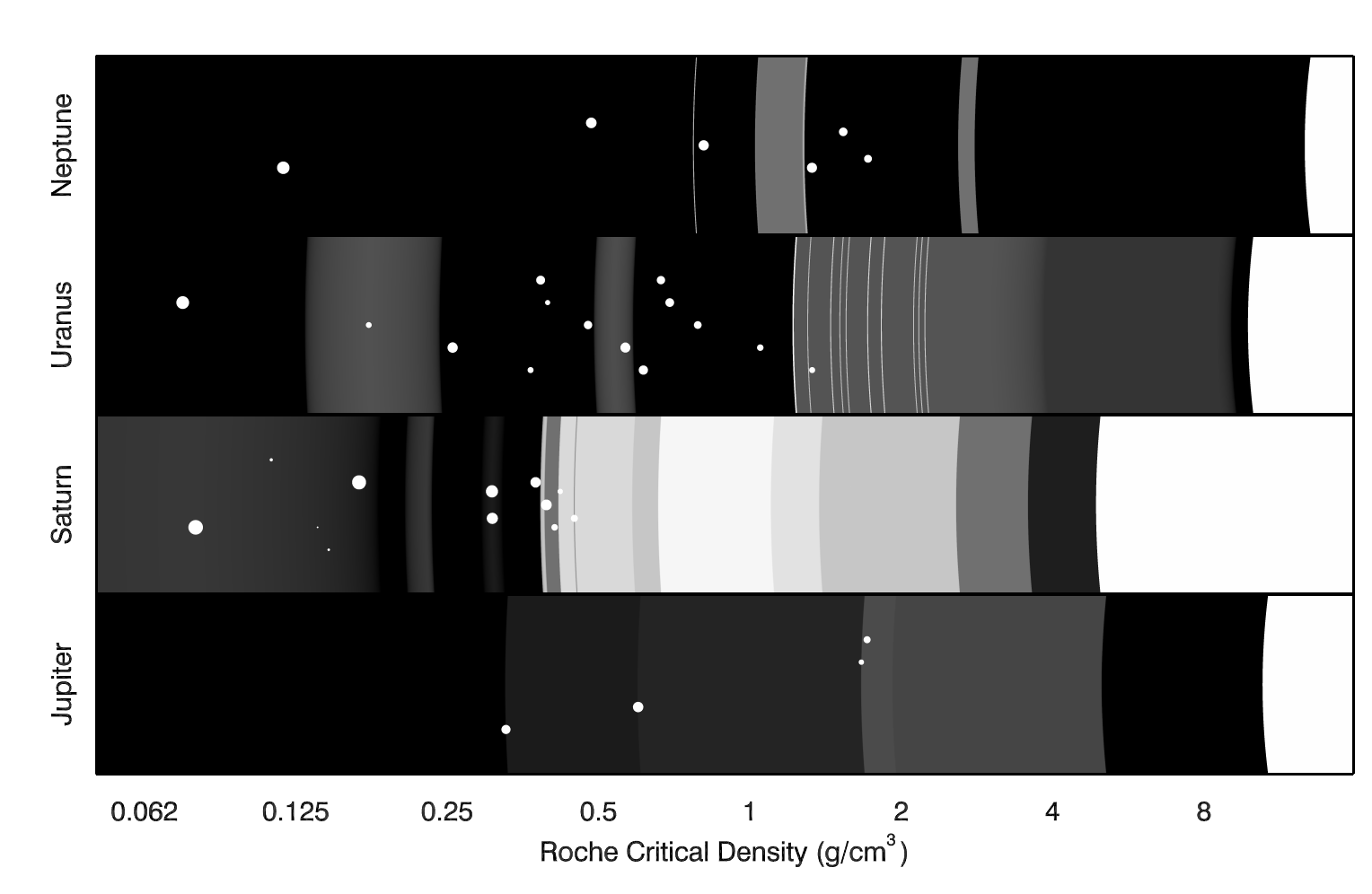}}
\caption{Schematic representations of the four giant planet systems, with the semi-major axes of the various rings and moons translated into Roche Critical Densities assuming $\gamma=1.6$. The shades of grey indicate the rings' optical depth (white being the most opaque), and the size of the spots indicate the relative sizes of the moons (not to scale with the rings).}
\label{RCD}
\end{figure}

 Consider a quantity known as  the ``Roche Critical Density'' \citep{Tiscareno13}:
\begin{equation}
\rho_{R}=\frac{3M_p}{\gamma a^3}.
\end{equation}
For a specified semi-major axis $a$, $\rho_{R}$ is the mass density an object would need to have in order for its Roche Limit $a_{R}=a$. Thus any isolated object with $\rho<\rho_{R}$ must have some internal strength. Figure~\ref{RCD} provides schematic illustrations of the ring-moon systems for the four giant planets, with the orbital semi-major axes of the various moons and rings translated into values of the Roche Critical Density.  Note that the innermost three moons of Neptune (Naiad, Thalassa and Despina) not only fall interior to most of Neptune's rings, they also occupy a region where the Roche Critical Density is comparable to that of Uranus' main ring system. This implies that Neptune's inner moons must have either larger mass densities or greater internal strengths than the material in Uranus' rings. The Jupiter system also has two moons (Metis and Adrastea) orbiting very close to the planet, where the Roche Critical Density exceeds 1.5 g/cm$^3$, again implying that these objects have high mass densities or significant strengths.

The mass densities of Jupiter's and Neptune's small inner moons are still not well constrained. However, it seems unlikely that  these objects would all  have sufficiently high mass densities to hold themselves together by gravity alone. The Roche Critical Densities at Metis' and Naiad's semi-major axes are about 1.7 g/cm$^3$. It would be somewhat unexpected if Naiad had such a large density, given that ices are expected to be the dominant solids around Neptune's orbit and Naiad is small enough to have a substantial porosity. Adrastea and Metis could potentially achieve suitably high densities if they were sufficiently compact and rock-rich, but the measured mass density of the nearby and larger moon Amalthea is only around  0.9 g/cm$^3$ \citep{Anderson05}. It would therefore be somewhat surprising for tiny Metis and Adrastea to have substantially higher mass densities than this. 


On the other hand, Naiad and Metis could plausibly have sufficient internal strength to hold themselves together. Over the years, various scholars have computed criteria under which objects with  finite tensile or shear strength should undergo failure due to tidal forces \citep{Davidsson99}. In general, the strength required for an object with density $\rho$ to survive at a location with a Roche Critical Density $\rho_{R}>\rho$ is given by an expression with the following form:
\begin{equation}
S=G\rho^2 R^2 \zeta(\rho_{R}/\rho),
\end{equation}
where $G$ is the universal gravitational constant, $R$ is  the object's radius, and $\zeta$ is a function of the ratio $\rho_{R}/\rho$ that is typically of order unity (The exact form of $\zeta$ depends upon whether the object splits in half or sheds material from its surface). Both Naiad and Metis have $R \simeq 30$ km, so assuming $\rho \simeq 1$ g/cm$^3$ and $\zeta \simeq 1$,  these objects would need to have an $S \simeq 6\times10^{4}$ N/m$^2$.While this is orders of magnitude greater than the estimated strengths of comets and rubble piles \citep{Greenberg95, Mohlmann95,SS14}, it is comparable to the strengths needed to support the observed topographies on small moons like Janus, Miranda or Phoebe \citep{Thomas13}. This is also well below the tensile strength of solid ice and is comparable to the strength of moderately dense snow \citep{Petrovic03}. Hence it is not unreasonable to suppose that Naiad and Metis are held together by their internal strength.

Similar calculations also place rather tight constraints on the strength of  Saturn's and Uranus' ring material. Thus far, the largest objects embedded in Saturn's rings (excluding moons like Pan that live near the Roche Limit) are less than 2 kilometers across \citep{Tiscareno10, Sremcevic14}. If we assume that the material in the rings has insufficient strength to hold together objects with $R>1$ km, then the strength of the ring material must be less than 100 N/m$^2$. Of course, this estimate of the ring material's strength neglects the role of collisions and other interactions among the ring particles in disrupting larger objects. However, the limited range of surface topographies observed on kilometer-sized comets and the smallest moons of Saturn also indicates that the regoliths on these bodies have effective strengths well below 100 N/m$^2$ \citep{Thomas13}. Aldditionally, this number  is comparable to the estimated strengths of rubble pile asteroids \citep{SS14}. Thus it is not unreasonable to suppose that the material in Saturn's and Uranus' rings is much weaker than the material in Jupiter's and Neptune's small moons. 

In fact, these sorts of considerations indicate that the material orbiting any body probably needs to be extremely weak to form dense rings. Uranus' epsilon ring  has a total mass of around $10^{16}$ kg \citep{French91}, which is comparable to the expected masses of moons like Metis and Naiad (assuming densities of order 1 g/cm$^3$). Hence if the material in Metis and Naiad was dispersed into a large number of meter-sized objects, then Jupiter and Neptune would likely have substantial rings. Conversely, if agglomerations of particles in Saturn's and Uranus' rings could  routinely achieve tensile and shear strengths  in excess of 100 N/m$^2$, then the material in these rings could accumulate into kilometer-scale objects and the rings' optical depth would be severely diminished. While there are not yet sufficient data to determine whether or not the material in the rings around Chariklo is similarly weak (e.g., Chariklo's total mass is not yet well constrained), Chariklo's ring material must also avoid aggregating together into larger particles for the rings to be detectable. Extremely weak orbiting material is therefore likely to be a necessary condition for the formation and maintenance of any dense rings.

\section{Dense rings have similar temperatures}

In order for the planet's distance to the Sun to influence its probability of having dense rings, the local solar flux must have some effect on the structure and/or dynamics of the relevant ring material. 
Solar radiation applies small forces to each ring particle (e.g. radiation pressure and Poynting-Robertson drag), and it also heats the ring material to a finite temperature. While solar perturbation forces could potentially be relevant to the orbital evolution of ring particles over the age of the Solar System, a comparison of the different rings' temperatures suggests that solar heating might be the more relevant phenomenon for the distribution of dense rings.

Extensive Cassini measurements show that the apparent temperature of Saturn's rings varies with  illumination and viewing geometries, as well as with radius across the rings \citep{Spilker13, Altobelli14, Filacchione14}. If we restrict our attention to the most massive parts of Saturn's rings (the A and B rings) these data indicate that the temperature of these regions ranges between 50  K and 90 K. However, this investigation is more concerned with the average effective temperature of this ring material, which can be roughly estimated using the standard expression derived from energy-balance considerations:
\begin{equation}
T_{\rm eff}=283 {\rm K}(1-A)^{1/4}\left({1 {\rm AU}}/{a_P}\right)^{1/2}
\end{equation}
where $A$ is the bolometric Bond albedo of the material and $a_P$ is the host planet's semi-major axis. This simple expression neglects heating from the planet itself and mutual shadowing among ring particles, but it should still  give a reasonable estimate of the ring's typical temperature. Indeed, \citet{Morishima10} estimated that $A= 0.5-0.7$ for Saturn's A and B rings, which yields $T_{\rm eff}= 66-75$ K, close to the average of the observed ring temperatures \citep{Altobelli14}. 

There are no direct temperature measurements of Uranus' or Chariklo's rings. However, the albedo of Uranus' rings is known to be only a few percent \citep{Ockert87, Karkoschka01}. This yields a $T_{\rm eff} \simeq 65$ K, which is remarkably close to the average temperature of Saturn's rings. Furthermore, Chariklo's rings appear to have an albedo that is intermediate between Saturn's and Uranus' rings \citep{Duffard14}, so the temperatures of these ring particles probably lie the same range.  It is a remarkable coincidence that three known  dense ring systems have very similar temperatures, so it is reasonable to ask whether the formation or maintenance of dense rings might be favored at temperatures close to 70 K.

\section{Is icy material particularly weak at 70 K?}

If we accept that dense rings need to be composed of very weak material to avoid aggregating into moons, then the restricted distribution of dense rings in our Solar System could arise if material only becomes suitably weak at temperatures around 70 K. On a coarse scale, it is reasonable to expect that material becomes weaker in the outer Solar System as ices become a larger fraction of the relevant solid matter. However, It is not yet clear whether these ices become particularly weak at temperatures close to 70 K.

One potential problem with the idea that icy materials becomes weak at a particular temperature is that most of Saturn's and Uranus' moons do not appear to be made of sufficiently weak material. The regolith on these moon's surfaces has to be strong enough to support topographic features like craters, and recent studies of Saturn's and Uranus' moons indicate that  Miranda, Janus and even Atlas can support topographic loads greater than 100 N/m$^2$ \citep{Thomas13}. Thus these moons seem to be much stronger than the ring material, despite their similar temperatures. However, the smallest moons of Saturn observed at high resolution by Cassini (Methone and Pallene) have very low topography, and like comets of comparable size, the supported topography requires strengths well below 100 N/m$^2$ \citep{Thomas13}, so these kilometer-wide objects in orbit around  Saturn do appear to be very weak. Thus it appears that only objects below a certain size have exceptionally low strengths. Most likely, this is because those objects have high porosities, which would naturally weaken them. 

Could porous aggregates be especially weak at temperatures close to 70 K? The topography of Methone and Pallene are well below what one would expect based on extrapolations from small asteroids like Steins and Itokawa \citep{Thomas13}, and so it does appear that small objects in orbit around Saturn are weaker than asteroids. Unfortunately, there are no similar high-resolution observations of kilometer-scale objects around other  giant planets, so we cannot yet directly determine whether Saturn's smallest moons are noticeably weaker than similar-sized objects orbiting Jupiter, Neptune or Kuiper Belt objects. However, there are some data which indirectly suggest that icy materials could be rather weak in the relevant parts of the outer Solar System.

The observed activity of comets and Centaurs in the outer Solar System  implies that some process allows volatiles to escape ice-rich bodies at relatively low temperatures \citep{Jewitt09}. Anything that would allow gases to escape from an active Centaur would probably also produce zones of weakness in the material, so perhaps the same basic mechanism responsible for the activity of Centaurs could be responsible for making the (porous) material in Saturn's and Uranus' rings sufficiently weak. One possible mechanism for weakening the material involves transformations between different phases of ice. The stability limits of amorphous ice and ice XI do both fall close to 70 K \citep{KK90, Arakawa11}. However, infrared spectra of Saturn's rings do not show clear evidence for amorphous ice, and spectra of Uranus' rings do not show any water-ice absorptions at all. It therefore appears unlikely that exotic ice phases are common in Uranus' or Saturn's rings, but it is not yet clear how much of this material would be needed to produce sufficiently weak icy bodies.

\section{Conclusions}

While we do not yet have a complete answer for why dense planetary rings are only found between 8AU and 20AU from the Sun, the above considerations do suggest that there might be some process that makes the icy material orbiting Saturn and Uranus exceptionally weak, enabling it to avoid agglomerating into moons. If icy materials naturally become weak at temperatures around 70 K,  then circumplanetary material will best be able to avoid accreting into moons when it is close to that temperature.  The albedos of Saturn's and Uranus' rings are close the maximum and minimum values found in planetary surfaces, and so it would be comparatively difficult for material to have this temperature outside of a restricted region in the outer Solar System. 

Fortunately, there are various avenues available to explore, test and refine these ideas. Most obviously, searches for rings around other small bodies could confirm or deny whether dense rings
are really restricted to the region around Saturn's and Uranus' orbits. Searches for rings around exoplanets could also be informative. For example, the above considerations would suggest that rings are unlikely to be found around  the close-in exoplanets that are most likely to produce transit signatures \citep{BF04, SC11}, but could be found further out, perhaps  as more compact versions of the disk that passed in front of the young star 1SWASP J140747.93-394542.6  \citep{Mamajek12, Kenworthy15, KM15}. Finally,  laboratory experiments could investigate whether ice-rich porous aggregates have unusual mechanical properties around 70 K, which is colder than the conditions explored by most of the experimental studies of icy materials \citep{Durham05, Choukroun12, Hill15a}.


\section*{Acknowledgements}

I thank M.S. Tiscareno, J.N. Spitale, P.D. Nicholson, J.A. Burns, P. Thomas, M. el Moutamid and D. Dhingra for useful conversations about these ideas. I also thank the anonymous reviewer for their helpful comments.

\end{document}